 \documentclass[preprint2]{aastex}
\usepackage{graphics,epsfig,longtable,lscape,txfonts,color}
\usepackage[english]{babel}
\newcommand{\myemail}{hstiele@mx.nthu.edu.tw}
\def\deg{\hbox{$^\circ$}}
\newcommand{\mr}{\mathrm}
\newcommand{\nh}{\hbox{$N_{\mr H}$}}
\newcommand{\hcm}[1]{$\times 10^{#1}$ cm$^{-2}$}
\def\subsun{\mbox{$_{\odot}$}}
\newcommand{\ergcm}[1]{$\times10^{#1}$~\hbox{erg~cm$^{-2}$~s$^{-1}$}}
\newcommand{\oergcm}[1]{$10^{#1}$~erg~cm$^{-2}$~s$^{-1}$}

\newcommand{\oergs}[1]{$10^{#1}$~erg~s$^{-1}$}

\def\ie{i.\,e.}                                      % i.e. (kursiv) \ie
\def\xmm{\textit{XMM-Newton}}
\def\swift{\textit{Swift}}
\def\nus{\textit{NuSTAR}}

\def\gx339{GX\,339-4}
\def\h1743{H\,1743-322}

\def\sw13{Swift\,J1357.2--0933}

\def\xu12{XMMU 122939.7+075333}

\shorttitle{2017 outburst of \sw13}
\shortauthors{Stiele, Kong}

\begin{document}

%% LaTeX will automatically break titles if they run longer than
%% one line. However, you may use \\ to force a line break if
%% you desire.

\title{\nus\ and \swift\ observations of \sw13\ during an early phase of its 2017 outburst}

\author{H.\ Stiele}
\affil{National Tsing Hua University, Department of Physics and Institute of Astronomy, No.~101 Sect.~2 Kuang-Fu Road,  30013, Hsinchu, Taiwan}
\email{\myemail}
\and

\author{A.\ K.\ H.\ Kong}
\affil{National Tsing Hua University, Department of Physics and Institute of Astronomy, No.~101 Sect.~2 Kuang-Fu Road,  30013, Hsinchu, Taiwan}
\affil{Astrophysics, Department of Physics, University of Oxford, Keble Road, Oxford OX1 3RH, UK}

%\date{Accepted 1988 December 15. Received 1988 December 14; in original form 1988 October 11}
%\date{2017 May 25}

%\pagerange{\pageref{firstpage}--\pageref{lastpage}} \pubyear{2017}

%\maketitle

%\label{firstpage}

\begin{abstract}
We present a detailed spectral analysis of \swift\ and \nus\ observations of the very faint X-ray transient and black hole system \sw13\ during an early low hard state of its 2017 outburst. \sw13\ was observed at $\sim$0.02\% of the Eddington luminosity (for a distance of 2.3 kpc and a mass of 4 M\subsun). Despite the low luminosity, the broadband X-ray spectrum between 0.3 and 78 keV requires the presence of a disk blackbody component with an inner disk temperature of $T_{\mr{in}}\sim0.06$ keV in addition to a thermal Comptonization component with a photon index of $\Gamma\sim1.70$. Using a more physical model, which takes strong relativistic effects into account, and assuming a high inclination of 70\deg, which is motivated by the presence of dips in optical light curves, we find that the accretion disk is truncated within a few $R_{\mr{ISCO}}$ from the black hole, independent of the spin. 
\end{abstract}


\section{Introduction}
\sw13\ is a black hole low-mass X-ray binary. It was first detected in 2011 by the \swift\ Burst Alert Telescope \citep{2005SSRv..120..143B,2011ATel.3138....1K}. The distance to the source is not well constrained and can range from $\sim$1.5 -- 6.3 kpc \citep{2011ATel.3140....1R,2013MNRAS.434.2696S}. If the true distance is towards the lower end of this range, \sw13\ belongs to the rare class of very faint X-ray transients \citep[VFXTs;][]{2006A&A...449.1117W}, which reach a peak luminosity of only $L_{\mr{X}}^{\mr{peak}}\sim10^{34}-10^{36}$ erg s$^{-1}$ \citep{2013MNRAS.428.3083A,2014MNRAS.439.3908A}. The mass of the black hole is estimated to be $>3.6$ M\subsun\ by \citet{2013Sci...339.1048C} or $>9.3$ M\subsun\ by \citet{2015MNRAS.454.2199M}.  An orbital period of $2.8\pm0.3$ h is obtained from time-resolved optical spectroscopy of broad, double-peaked H$\alpha$ emission \citep{2013Sci...339.1048C}. In the optical light curve recurring dips on 2--8 min time-scales have been discovered \citep{2013Sci...339.1048C}. In these dips the flux dropped by up to $\sim$0.8 mag. This short-term variability has been addressed to \sw13\ being viewed at a nearly edge-on inclination angle ($i\ga70$\deg) and the presence of a geometrically thick obscuring torus in the inner disk \citep{2013Sci...339.1048C}. The observed broad, double-peaked H$\alpha$ profile supports a high orbital inclination \citep{2015MNRAS.450.4292T}. The \textit{RXTE} and \xmm\ data do neither show signs of the 2.8 h orbital period, nor of the recurring dips seen in the optical \citep{2014MNRAS.439.3908A}.  

Renewed activity of \sw13\ was first reported as an optical outburst on 2017 April 20 by \citet{2017ATel10297....1D}. Subsequently \swift/XRT observation taken on April 21 detected a bright X-ray source at the optical position \citep{2017ATel10314....1S} suggesting an early phase of an X-ray outburst. This provides a good opportunity to study a black hole X-ray transient during its rising outburst. We therefore requested for a \nus\ target of opportunity (ToO) observation, to observe \sw13\ during an early phase of its outburst. There are also two \swift/XRT observations taken simultaneously to the \nus\ observation.

\section[]{Observation and data analysis}
\label{Sec:obs}
\subsection{\nus}
We analysed the \nus\ \citep{2013ApJ...770..103H} data taken on April 28 (obsid: 90201057002; exposure: 34.5 ks) using the NuSTARDAS tools \texttt{nupipeline} and \texttt{nuproducts}. To extract source photons we used a circular region with a radius of 30\arcsec\ located at the known position of \sw13. Photons for the background spectra were extracted from a region of the same shape and size located close to the source on the same detector that was free of source photons. To investigate short term variability we derived cospectra in the 3 -- 30 keV band using MaLTPyNT \citep{2015ascl.soft02021B}. The energy range between 3 and 30 keV comprises about  97.3 per cent of the source photons detected with \nus\ in the 3 -- 78 keV band. The cospectrum is the cross power density spectrum (PDS) derived from data of the two completely independent focal planes and represents a good proxy of the white-noise-subtracted PDS \citep{2015ApJ...800..109B}. 

\subsection{\swift} 
We analysed \swift/XRT \citep{2005SSRv..120..165B} monitoring data, using the online data analysis tools provided by the Leicester \swift\ data centre\footnote{http://www.swift.ac.uk/user\_objects/}, including single pixel events only \citep{2009MNRAS.397.1177E}. Two observations were taken simultaneously to our \nus\ ToO observation: the first observation (obsid: 00088094002; PI: Stiele) was taken in photon counting mode, while the second observation (obsid: 00031918053; PI: Sivakoff) was taken in windowed timing mode. In the first observation the exposure was split in two parts of 196 s and 765s, with a gap of 5615 s in between. The exposure of the second observation was 524s.

For the \swift/UVOT \citep{2005SSRv..120...95R} data, we used the task \texttt{uvot2pha} to extract spectral information. In the first observation images in the U and UVW1 filter are taken, while in the second observation B, V, U, UVW1, UVW2, and UVM2 images are available.
 
\section[]{Results}
\label{Sec:res}

\subsection{Timing analysis}
\begin{figure}
\centering
\resizebox{\hsize}{!}{\includegraphics[clip,angle=0]{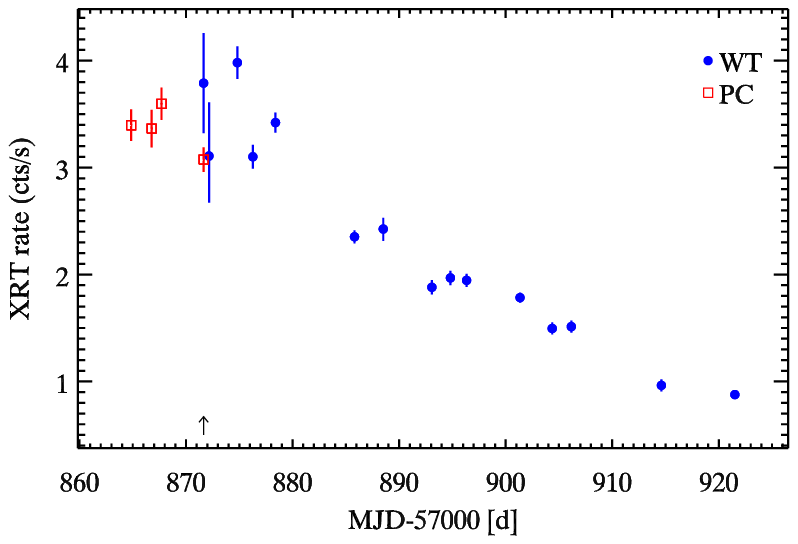}}
\caption{\swift/XRT light curve  of the 2017 outburst. Observations taken in photon counting mode are indicated as (red) open squares, those taken in windowed timing mode as (blue) open circles. The arrow indicates the time of our \nus\ observation.}
\label{Fig:lc_sw}
\end{figure}

The \swift/XRT longterm light curve is shown in Fig\ \ref{Fig:lc_sw}. The \nus\ observation was taken near the outburst peak. 

\begin{figure*}
\centering
\resizebox{\hsize}{!}{\includegraphics[clip,angle=0]{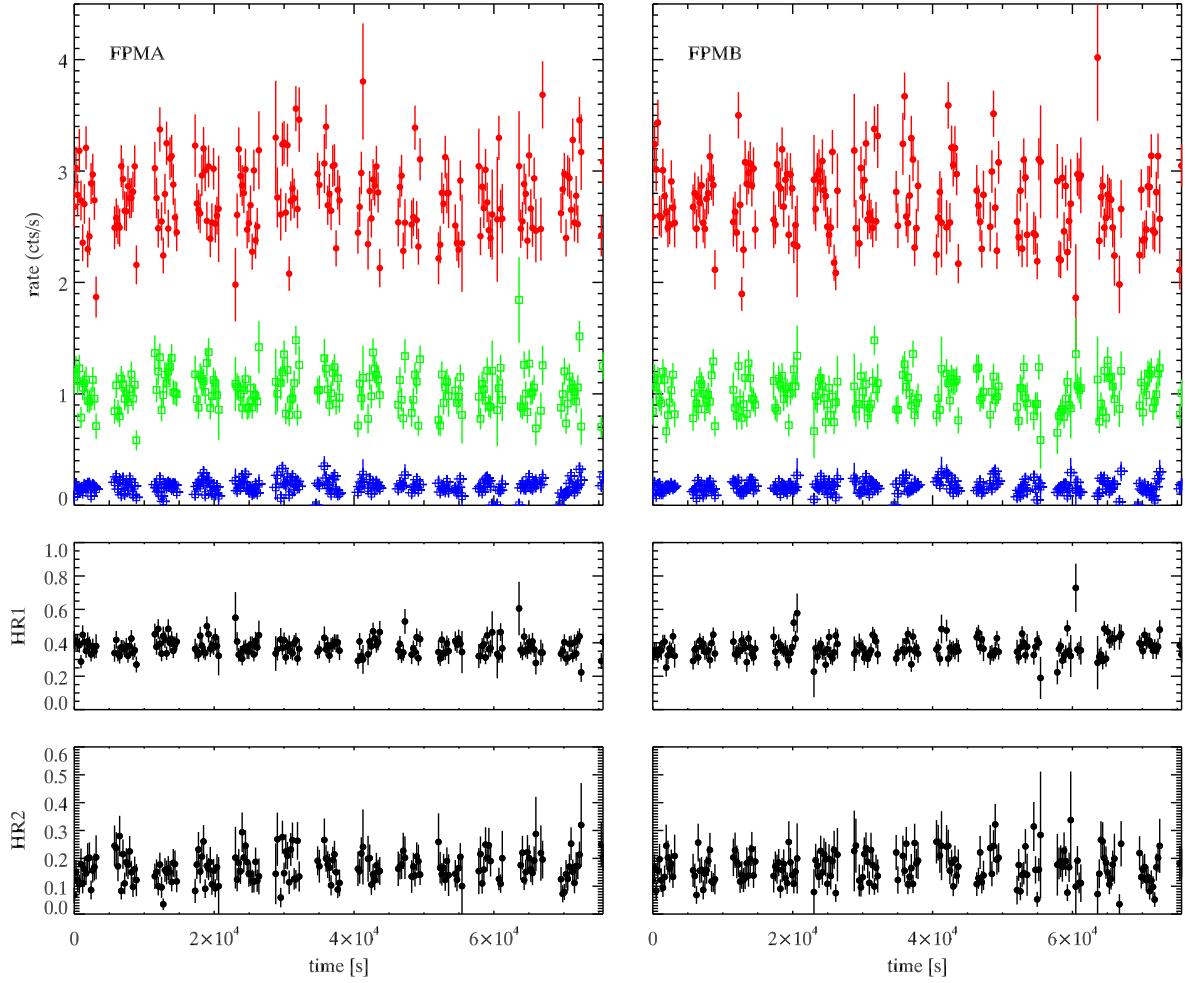}}
\caption{\nus\ FPMA and FPMB light curves (240 s bins) in the low 3 -- 10 (red filled dots), medium 10 -- 25 (green open squares), and high 25 -- 78 keV (blue crosses) bands. In the lower panels hardness ratios are shown (HR1=L/M; HR2=M/H).}
\label{Fig:lc_nus}
\end{figure*}

\begin{figure}
\centering
\resizebox{\hsize}{!}{\includegraphics[clip,angle=0]{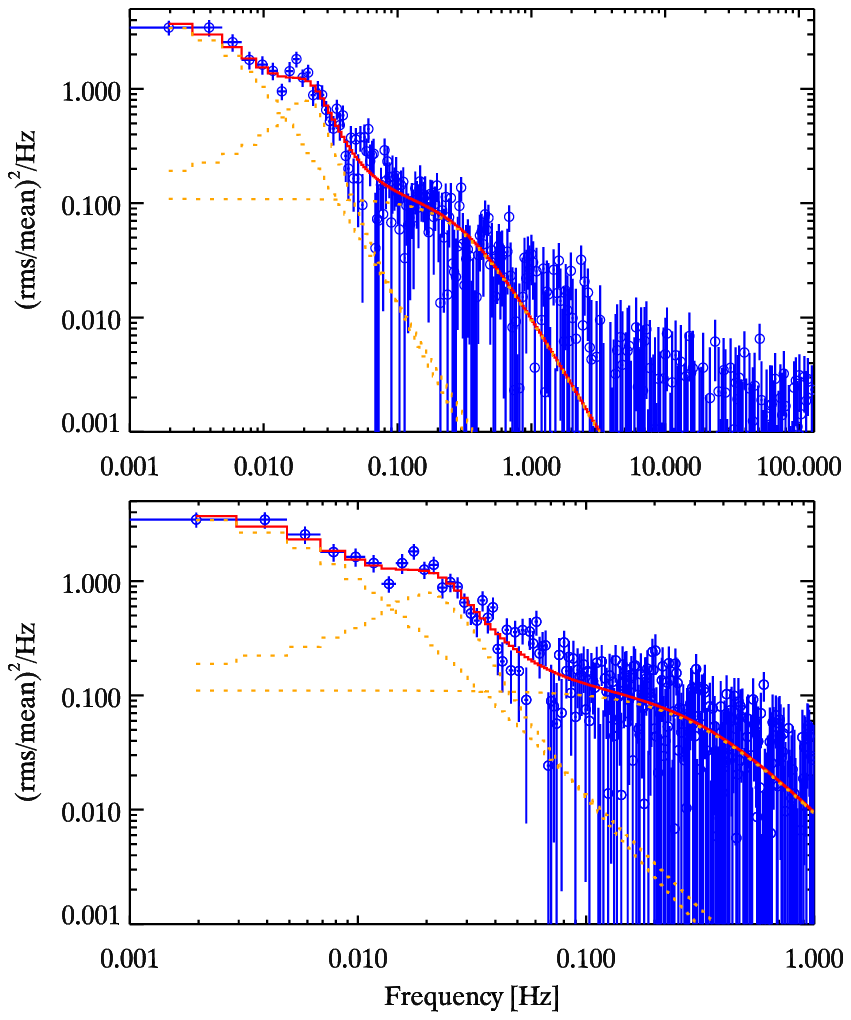}}
\caption{Cospectra in the $\sim2^{-3}$ -- 128 Hz (upper panel) and $\sim2^{-3}$ -- 1 Hz (lower panel) range.}
\label{Fig:pds}
\end{figure}

For the \nus\ data, light curves in three energy bands with a binning of 240 s, and corresponding hardness ratios are shown in Fig.\ \ref{Fig:lc_nus}. We derived cospectra in the 3 -- 30 keV range, using time bins of $2^{-8}$ s and of 0.5 s and stretches of 512 s, corresponding to frequencies between $\sim2^{-3}$ Hz and 128 or 1 Hz, respectively (Fig.~\ref{Fig:pds}). The cospectra can be fitted with two zero-centred Lorentzians, modelling band-limited noise (BLN) components, and one Lorentzian to fit a peaked noise (PN) component. The BLN components have an rms variability of $18.9^{+1.6}_{-1.4}$\% and $23.0^{+0.9}_{-1.0}$\%, respectively, and a characteristic frequency of $\nu=6.0^{+1.6}_{-1.1}$ mHz and $\nu=0.31^{+0.06}_{-0.05}$ Hz, respectively. The PN component has a centroid frequency of $\nu_0=20.3\pm1.8$ mHz, a half width at half maximum of $\Delta=10.3^{+2.7}_{-2.8}$ mHz, an rms variability of $14.8^{+1.8}_{-2.2}$\%, and a significance of 3.3$\sigma$. The parameters are almost identical in both bands. 

\subsection{Spectral analysis}
\label{Sec:spec}
\begin{figure*}
\centering
\resizebox{0.8\hsize}{!}{\includegraphics[clip,angle=0]{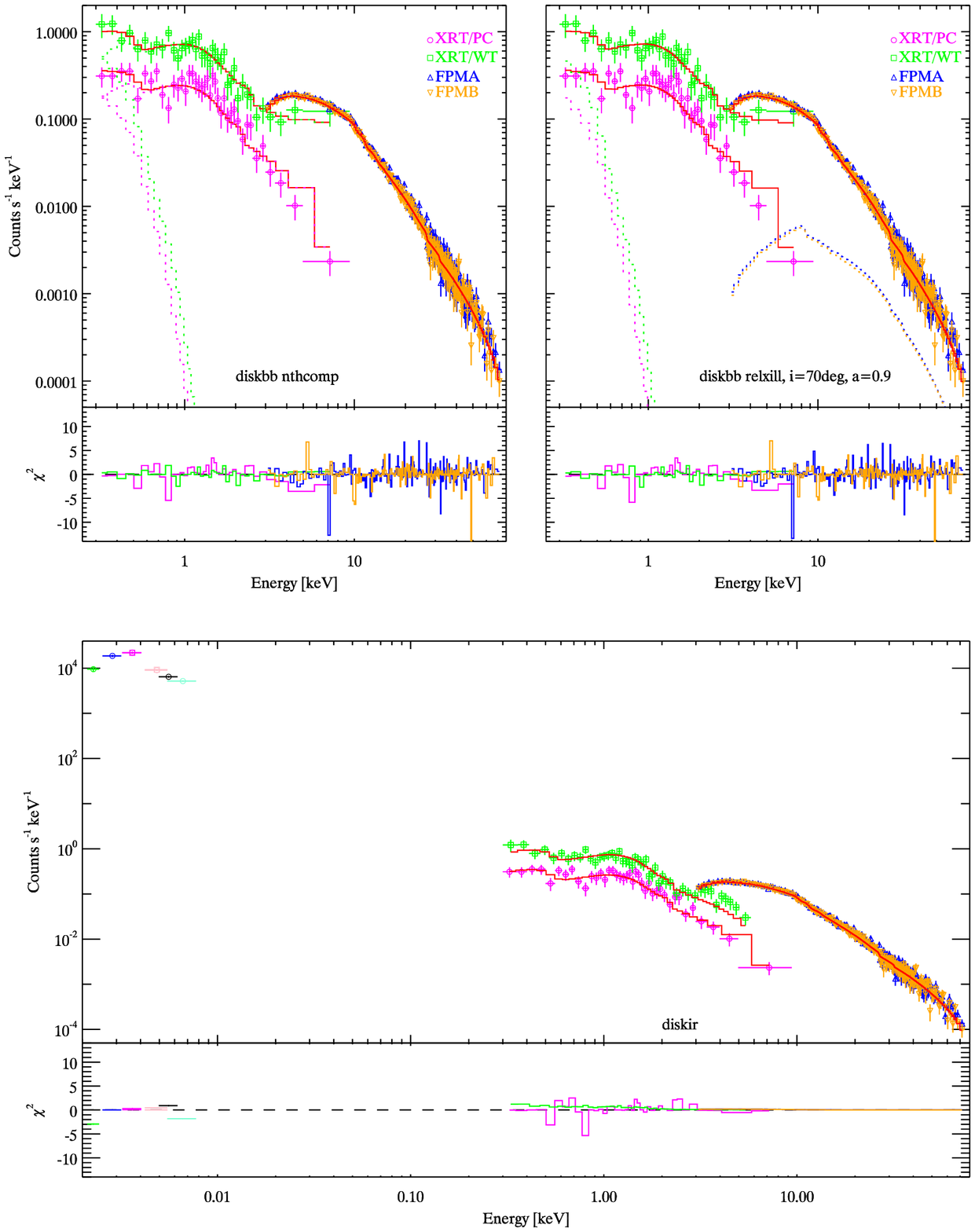}}
\caption{\swift/XRT and \nus\ energy spectra fitted with an absorbed \texttt{diskbb} + \texttt{nthcomp} (upper left panel) and \texttt{diskbb} + \texttt{relxill} model, with $i=70$\deg\ and $a=0.9$ (upper right panel). The disk component and the reflection component are indicated by dashed lines. The lower panel shows the \swift/UVOT, XRT and \nus\ energy spectra fitted with the \texttt{diskir} model.}
\label{Fig:spec}
\end{figure*}

We fit the averaged energy spectra of the \nus\ and \swift\ ToO observations within \textsc{isis} \citep[V.~1.6.2;][]{2000ASPC..216..591H} in the 0.3 -- 78 keV range, where the \swift/XRT data cover the 0.3 -- 10 keV range and the \nus\ data cover the 3 -- 78 keV range. We grouped the \swift/XRT data to a signal to noise ratio of 3 and the \nus\ data to a signal to noise ratio of 5, with a least five channels per bin, both for \swift/XRT and \nus. The \swift/XRT and \nus\ spectra can be well fitted with an absorbed power law model ($\chi^2/dof$: 539.7/525; Fig.\ \ref{Fig:spec}), where the absorption, modelled with \texttt{TBabs} \citep{2000ApJ...542..914W}, is fixed at  \nh = 3\hcm{20}. We use the abundances of \citet{2000ApJ...542..914W} and the cross sections given in \citet{1996ApJ...465..487V}. The obtained photon index is $\Gamma=1.679\pm0.006$. The spectral parameters are given in Table \ref{Tab:spec_par}. We also add a floating cross-normalisation parameter, which is fixed to one for \nus\ FPMA, to take uncertainties in the cross-calibration between the different telescopes into account. The absorbed flux in the 0.3 -- 78 keV band is $2.08\pm0.08$\ergcm{-9}, and the unabsorbed flux in the 0.5 -- 10 keV band is $2.75\pm0.11$\ergcm{-10}, a factor $\sim$2.6 bigger than the unabsorbed flux observed on April 21 \citep{2017ATel10314....1S}. The uncertainties on the flux are obtained following the approach presented in \citet{2004ApJ...606L..61W}.

We also try to fit a cut-off power law to the \swift\ and \nus\ spectra. This model gives a cut-off energy $\ga428$ keV, outside the energy range covered by the data. The values of all other parameters are similar to the one found using the power law model (see Table \ref{Tab:spec_par}). The cut-off power law is not statistically required compared with the simple power law model ($\chi^2$/dof: 541.3/524). 

Using a broken power law (\texttt{bknpower}), we obtain a statistically improved fit ($\chi^2/dof$: 526.1/523). According to the sample-corrected Akaike Information Criterion \citep[AIC][]{1974ITAC...19..716A}, this is an improvement of $\Delta$AIC$=9.5$, \ie, a 115 times better model \citep{burnham2011}. The spectral parameters and the flux is given in Table \ref{Tab:spec_par}.

As the broken power law provides a much better fit than the simple power law, we also fit the spectra with an absorbed disk blackbody (\texttt{diskbb}) plus thermal Comptonization model \citep[\texttt{nthcomp}][]{1996MNRAS.283..193Z,1999MNRAS.309..561Z}. This results in a further improved fit of $\chi^2/dof$: 520.3/522. The spectral parameters and the flux can be found in Table \ref{Tab:spec_par}. To investigate a possible dependency of the disk parameters on the absorption, which is low \citep{2014MNRAS.439.3908A,2016MNRAS.456.2707P}, we fit the spectra with rather extreme absorptions of \nh = 0 and 1\hcm{21}. We find that with increasing absorption the disk temperature increases, while the disk radius decreases. However, these changes are small and the spectral parameters obtained with these two absorption values are consistent within errors.

\begin{figure}
\centering
\resizebox{\hsize}{!}{\includegraphics[clip,angle=0]{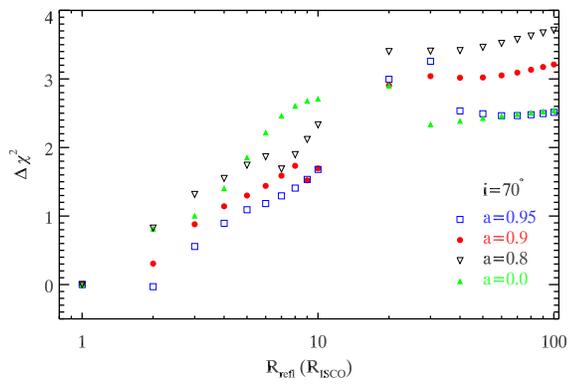}}
\caption{Change of $\chi^2$ with different values of the disk radius seen by the reflection component.}
\label{Fig:rin_chi}
\end{figure}

\begin{table*}
\caption{Averaged \swift\ and \nus\ spectra of \sw13\ fitted with different phenomenological models.}
\begin{center}
\small
\begin{tabular}{llllllll}
\hline\noalign{\smallskip}
 & \multicolumn{4}{c}{\swift/XRT \& \nus}  & \multicolumn{2}{c}{+ \swift/UVOT} \\
\hline\noalign{\smallskip}
 \multicolumn{1}{c}{parameter} & \multicolumn{1}{c}{powerlaw}  & \multicolumn{1}{c}{cutoffpl} & \multicolumn{1}{c}{bknpower} & \multicolumn{1}{c}{diskbb+nthcomp} & \multicolumn{1}{c}{diskbb+nthcomp} & \multicolumn{1}{c}{diskir} \\
 \hline\noalign{\smallskip}
$\Gamma$ &  $1.679\pm0.006 $   &  $1.656\pm0.006$ &  $1.678^{+0.005}_{-0.000}$ & $1.703^{+0.004}_{-0.005}$ & $1.702\pm0.005$&$1.697\pm0.005$\\            
\smallskip
$\Gamma_2$ & -- & -- & $2.46^{+1.83}_{-0.26}$&--&--&--\\
\smallskip
norm $^*\times10^{-2}$& $2.47\pm0.03 $  &  $2.40\pm0.03 $ &  $2.01^{+0.30}_{-1.60}$ & $2.21_{-1.62}^{+0.22} $ & $2.51\pm0.02$&--\\           
\smallskip
E$_{\mr{cut/break/e}}$ (keV) & -- &  $\ga427.7$& $0.77^{+0.12}_{-0.27}$& $47.2_{-12.7}^{+40.5}$ & $46.0_{-11.3}^{+41.1}$&$32.9_{-4.5}^{+9.1}$\\           
\smallskip
R$_{\mr{in}}^{\ddagger}$ (km)  &--& -- & -- & $367.67_{-266.53}^{+1191.87}$ &  $1395.33_{-108.65}^{+103.37}$ & $108.87_{-25.81}^{+188.65}$\\           
\smallskip
T$_{\mr{in}}$ (eV)  &--& --& --& $47.2_{-12.7}^{+40.5}$ &  $43.5_{-1.9}^{+2.1}$&$97.5_{-14.7}^{+10.8}$\\           
\smallskip
$L_{\mr{c}}/L_{\mr{d}}$& -- & -- & --&--&--&$3.02_{-1.01}^{+0.86}$\\
\smallskip
$R_{\mr{irr}}$ (R$_{\mr{in}}$)& -- & -- & --&--& --&$1.003\pm0.001$\\
\smallskip
$f_{\mr{out}}$ & -- & -- & --&--& --&$0.0122_{-0.0022}^{+0.0055}$\\
\smallskip
$R_{\mr{out}}$ (log$_{10}$R$_{\mr{in}}$)& -- & -- & --&--&--&$3.48_{-0.21}^{+0.04}$\\
\smallskip
$E(B-V)$ (mag) & -- & -- & --&--& $0.075\pm0.010$&$<0.042$\\
\smallskip
const$_{\mr{FPMB}}$ & $0.992\pm0.006 $   &  $0.992\pm0.006 $&  $0.992\pm0.006 $& $0.992\pm0.006 $  & $0.992\pm0.006$ & $0.992\pm0.006$\\         
\smallskip
const$_{\mr{XRT}}$ & $0.541^{+0.030}_{-0.029} $  &  $0.584\pm0.029 $&  $0.541^{+0.030}_{-0.029} $&$0.539^{+0.029}_{-0.028} $&$0.545^{+0.028}_{-0.029}$&$0.417^{+0.024}_{-0.025}$\\         
\smallskip
const$_{\mr{XRT2}}$ & $0.627^{+0.032}_{-0.028} $  & $0.660_{-0.028}^{+0.029} $ & $0.627^{+0.032}_{-0.028}$ &$0.625^{+0.028}_{-0.026} $&$0.631\pm0.027$&$0.455\pm0.023$\\         
\smallskip
$\chi^2/dof$ & 539.7/525 & 541.3/524& 526.1/523 & 520.3/522 & 557.3/521 & 526.4/518\\
\smallskip
F$^{\dagger}$ &$2.08\pm0.08$& $2.14\pm0.04$ & $2.54^{+0.28}_{-0.12}$&$2.07\pm0.08$&--&--\\
\hline\noalign{\smallskip} 
\end{tabular} 
\end{center}
Notes:\\
$^{\dagger}$: absorbed flux in the 0.3 -- 78 keV band in units of \oergcm{-9} \\
$^{\ddagger}$: assuming an inclination of 70\deg\ and a distance of 2.3 kpc\\
$^{*}$: in units of photons/keV/cm$^2$/s at 1 keV
\label{Tab:spec_par}
\end{table*}

While the phenomenological models presented above provide a statistically very good fit, they do not contain information about the geometry of the X-ray producing region. To obtain information about the geometry we need to study the strong relativistic effects close to the black hole. We model the relativistic effects using the \texttt{relxill} model \citep{2014MNRAS.444L.100D,2014ApJ...782...76G}. We use an emissivity index of 3, which is appropriate for a standard Shakura-Sunyaev accretion disk and an extended corona \citep{1997ApJ...488..109R}. Due to the low count rates and the degeneracy of different spectral parameters of the model, we need to fix some values. As the cut-off power law model gives a high cut-off energy we keep the cut-off energy in the \texttt{relxill} model at 300 keV. We also set the outer disk radius to $r_{\mr{out}}=400 R_g$. We try two different inclinations of 30\deg\ and 70\deg, where the higher inclination is motivated by dip-like features observed in optical light curves. As there are no measurements of the spin available from the literature, we try values of zero (Schwarzschild black hole), 0.8, 0.9, and 0.95 for the spin.

Using the higher inclination we obtain a photon index of $\Gamma\sim1.70$, an inner disc temperature of $T_{\mr{in}}\sim56$ eV and a low reflection of $\mathcal{R}_{\mr{refl}}\sim0.16$. The iron abundance (A$_{\mr{Fe}}<0.8$), and the ionisation parameter (log($\xi/$(erg cm s$^{-1}$)) $\sim1.0$) are not well constrained. That the presence of a disk component hampers constraining the ionisation parameter has already been noticed in previous studies \citep{2016MNRAS.458.2199B,2017ApJ...844....8S}. The values of the spectral parameters are consistent for the different spin values (see Table \ref{Tab:spec_par2}). The inner radius of the reflection disk (measured in $R_{\mr{ISCO}}$) increases with increasing spin, but the data prefer an accretion disk truncated close (within $\la5R_{\mr{ISCO}}$) to the black hole. We show in Fig.~\ref{Fig:rin_chi} how $\chi^2$ changes with the inner disk radius compared to the best-fit $\chi^2$ value. These relations confirm that a disk truncated close to the black hole is preferred, independent of the spin value. The flux of the direct disk component contributes about one third of the total flux in the 0.3 -- 1 keV band.     
For the lower inclination of 30\deg, which gives statistically slightly worse fits with $\Delta\chi\sim+1-2$ (for the same dof), we obtain a photon index of  $\Gamma\sim1.68$, an inner disc temperature of $T_{\mr{in}}\sim59$ eV and an even lower reflection fraction of $\mathcal{R}_{\mr{refl}}\sim0.11$, independent of the spin. The iron abundance (A$_{\mr{Fe}}<0.9$) and the ionisation parameter (log($\xi/$(erg cm s$^{-1}$)) $\sim2.0$) are not well constrained. The inner disk radius is also not well constrained, but now the fits prefer a bigger inner disk radius with a disk truncated above $\sim3R_{\mr{ISCO}}$.    

Assuming a lamppost geometry and an inclination of 70\deg, the obtained spectral parameters are in agreement with those given in Table \ref{Tab:spec_par2} for the \texttt{relxill} model. For the disk height we find upper limits of 20.01, 15.47, 18.53, and 19.47 $R_g$ for spin values of 0, 0.8, 0.9, and 0.95, respectively. For an inclination of 30\deg, the inner disk radius is unconstrained and we obtain lower limits on the height of the accretion disk of 5.78, 6.18, and 6.34 $R_g$ for spin values of 0.8, 0.9, and 0.95, respectively. For a Schwarzschild black hole the disk height is unconstrained. 

\begin{table*}
\caption{Averaged \swift/XRT and \nus\ spectra of \sw13\ fitted with an absorbed relxill model assuming different inclination and spin.}
\begin{center}
\begin{tabular}{lllll}
\hline\noalign{\smallskip}
 \multicolumn{1}{c}{parameter} & \multicolumn{1}{c}{$a=0$}  & \multicolumn{1}{c}{$a=0.8$} & \multicolumn{1}{c}{$a=0.9$}& \multicolumn{1}{c}{$a=0.95$}\\
 \hline\noalign{\smallskip}
& \multicolumn{4}{c}{Inclination: 70\deg} \\
 \hline\noalign{\smallskip}
R$_{\mr{in}}^{\ddagger} $ (km)  &$452.64_{-357.10}^{+2482.27}$& $461.33_{-366.55}^{+2262.94}$ & $464.98_{-370.53}^{+2140.77}$ & $441.37_{-346.39}^{+2232.74}$\\           
\smallskip
T$_{\mr{in}}$ (eV)  &$56.0_{-18.7}^{+29.8}$& $55.8_{-18.0}^{+30.0}$ & $55.7_{-17.4}^{+30.1}$ & $56.4_{-18.4}^{+29.4}$\\           
\smallskip
$\Gamma$ &  $1.70\pm0.02 $  &  $1.70_{-0.03}^{+0.02}$ &  $1.70\pm0.02$  & $1.70\pm0.02$\\            
\smallskip
norm $\times10^{-2}$& $2.55_{-0.14}^{+0.08} $ &  $2.55\pm0.08 $ &  $2.55_{-0.18}^{+0.08} $  & $2.55_{-0.18}^{+0.08}$ \\           
\smallskip
R$_{\mr{refl}} (R_{\mr{ISCO}})$  &$<2.75$& $<2.26$ & $<3.49$ & $<5.72$\\           
\smallskip
log$\xi$ (erg cm s$^{-1}$)  & $0.92_{-0.92}^{+1.61}$ & $0.95_{-0.95}^{+1.79}$ & $0.99_{-0.99}^{+1.74}$ & $1.00_{-1.00}^{+1.74}$\\           
\smallskip
A$_{\mr{Fe}}$  & $<0.81$ & $<0.81$ & $<0.79$ & $<0.80$\\           
\smallskip
$\mathcal{R}_{\mr{refl}}$  & $0.17\pm0.06$ & $0.17\pm0.06$ & $0.16_{-0.05}^{+0.06}$ & $0.16\pm0.06$\\           
\smallskip
const$_{\mr{FPMB}}$ & $0.992\pm0.006 $ &  $0.992\pm0.006 $  &  $0.992\pm0.006 $&  $0.992\pm0.006 $\\         
\smallskip
const$_{\mr{XRT/PC}}$ & $0.533_{-0.030}^{+0.031} $  &  $0.532_{-0.030}^{+0.031} $ &  $0.533_{-0.031}^{+0.030} $&  $0.533_{-0.031}^{+0.030} $\\         
\smallskip
const$_{\mr{XRT/WT}}$ & $0.617\pm0.030 $   & $0.616_{-0.033}^{+0.030} $  & $0.617_{-0.031}^{+0.030} $  & $0.617_{-0.035}^{+0.030} $  \\         
\smallskip
$\chi^2/dof$ &  517.7/519& 517.2/519 & 517.4/519&517.5/519 \\
\smallskip
F$^{\dagger}$ & $2.08\pm0.06$ & $2.08\pm0.06$&$2.07\pm0.06$ & $2.07\pm0.06$\\
\hline\noalign{\smallskip} 
& \multicolumn{4}{c}{Inclination: 30\deg} \\
 \hline\noalign{\smallskip}
R$_{\mr{in}}^{\ddagger}$ (km)  &$235.84_{-171.07}^{+253.62}$& $238.48_{-174.05}^{+1149.38}$ & $233.67_{-169.22}^{+160.65}$ & $220.49_{-156.28}^{+1176.39}$\\           
\smallskip
T$_{\mr{in}}$ (eV)  &$58.9_{-17.5}^{+23.1}$& $58.7_{-17.3}^{+28.2}$ & $59.0_{-19.4}^{+27.9}$ & $59.9_{-20.4}^{+27.2}$\\           
\smallskip
$\Gamma$ &  $1.68_{-0.08}^{+0.02}$  &  $1.68\pm0.02$ &  $1.68\pm0.02$  & $1.68\pm0.02$ \\            
\smallskip
norm $\times10^{-2}$& $2.48\pm0.06 $ &  $2.48_{-0.04}^{+0.06} $ &  $2.48_{-0.04}^{+0.06} $  & $2.48_{-0.06}^{+0.05}$  \\           
\smallskip
R$_{\mr{refl}} (R_{\mr{ISCO}})$  &$>1.00$& $>1.73$ & $>2.16$ & $>2.66$ \\           
\smallskip
log$\xi$  (erg cm s$^{-1}$) & $1.98_{-1.98}^{+0.43}$ & $2.00_{-2.00}^{+0.42}$ & $2.00_{-2.00}^{+0.42}$ & $1.99_{-1.99}^{+0.43}$ \\           
\smallskip
A$_{\mr{Fe}}$  & $<0.94$ & $<0.94$ & $<0.93$ & $<0.92$ \\           
\smallskip
$\mathcal{R}_{\mr{refl}}$  & $0.11\pm0.04$ & $0.11\pm0.04$ & $0.11\pm0.04$ & $0.11\pm0.04$ \\           
\smallskip
const$_{\mr{FPMB}}$ & $0.992\pm0.006 $ &  $0.992\pm0.006 $  &  $0.992\pm0.006 $&  $0.992\pm0.006 $  \\         
\smallskip
const$_{\mr{XRT/PC}}$ & $0.538_{-0.029}^{+0.031} $  &  $0.538_{-0.029}^{+0.030} $ &  $0.537_{-0.029}^{+0.031}$&  $0.537_{-0.029}^{+0.030}$  \\         
\smallskip
const$_{\mr{XRT/WT}}$ & $0.624_{-0.028}^{+0.030} $   & $0.624_{-0.020}^{+0.030} $  & $0.623_{-0.028}^{+0.030} $  & $0.623_{-0.020}^{+0.029} $  \\         
\smallskip
$\chi^2/dof$ & 519.0/519 & 519.0/519& 519.0/519& 519.0/519\\
\smallskip
F$^{\dagger}$& $2.06^{+0.06}_{-0.08}$ & $2.06^{+0.06}_{-0.09}$& $2.06^{+0.06}_{-0.08}$& $2.06^{+0.06}_{-0.08}$\\
\hline\noalign{\smallskip} 
\end{tabular} 
\end{center}
Notes:\\
$^{\dagger}$: absorbed flux in the 0.3 -- 78 keV band in units of \oergcm{-9} \\
$^{\ddagger}$: assuming a distance of 2.3 kpc
\label{Tab:spec_par2}
\end{table*}

Adding the \swift/UVOT data points and using the absorbed power law and broken power law model, we obtain photon indices and a break energy consistent with those of the \swift/XRT and \nus\ fit with the same model given in Table \ref{Tab:spec_par}. Fitting just the \swift/UVOT data points with a power law model results in a statistically unacceptable fit with a photon index inconsistent with $\Gamma\sim1.6-1.7$, which confirms a change of the spectral shape at soft energies. Fitting the combined \swift/UVOT, XRT and \nus\ spectra with an absorbed disk blackbody plus thermal Comptonization model, where we used the \textsc{redden} component \citep{1989ApJ...345..245C} for the interstellar extinction in the UVOT filters, we obtain spectral parameters consistent within errors with those of the \swift/XRT and \nus\ fit with the same model (see Table \ref{Tab:spec_par}).  
We also fit the spectra with the \textsc{diskir} model \citep{2008MNRAS.388..753G,2009MNRAS.392.1106G}, which takes irradiation of the accretion disk into account. The obtained parameters can be found in Table \ref{Tab:spec_par}. Including disk irradiation gives a higher disk temperature and a smaller disk radius, but the obtained parameters are consistent within errors with those of the \swift/XRT and \nus\ fit with the absorbed disk blackbody plus thermal Comptonization model. 
\citet{2013MNRAS.428.3083A} found that the correlation between \swift/UVOT v-band and XRT data is consistent with a non-irradiated accretion disk. The data point obtained from the 2017 outburst lies below the correlation shown in \citet{2013MNRAS.428.3083A}. As we only have this datapoint the correlation between v-band and XRT flux does not help us to decide if disk irradiation plays a role in the 2017 outburst.  The obtained interstellar extinction is at least as big as the Galactic reddening in the direction of \sw13\ \citep[$E(B-V)=0.04$ mag;][]{1998ApJ...500..525S}.

\section[]{Discussion}
\label{Sec:dis}
We have investigated the temporal and spectral properties of \sw13\ observed with \swift/XRT and \nus\ during an early phase of its 2017 outburst. The combined \swift\ and \nus\ spectra can be fitted well with an absorbed power law model with a photon index of $1.679\pm0.006$, which agrees with the value of $1.69\pm0.13$ obtained from \swift/XRT data at a similar 0.5 -- 10 keV luminosity during the 2011 outburst  \citep{2013MNRAS.428.3083A}. A statistically even better fit can be obtained using an absorbed disk blackbody plus thermal Comptonization model with a photon index of $1.703_{-0.005}^{+0.004}$ and an inner disk temperature of $59.7_{-15.1}^{+26.8}$ eV. This is a much lower temperature and a softer photon index than the one observed during the 2011 outburst with \xmm, at an about 1.2 times higher unabsorbed 0.5 -- 10 keV band flux \citep{2014MNRAS.439.3908A}.

While these phenomenological models provide statistically good fits, they do not allow us to learn anything about the accretion geometry of the system. As we need to study the strong gravity effects close to the black hole to obtain information about the geometry, we fitted the spectra with an absorbed disk blackbody plus \texttt{relxill} model. Assuming a high inclination of 70\deg, which is motivated by the observed dips in optical light curves \citep{2013Sci...339.1048C}, we find that the disk is truncated close to the black hole within a few $R_{\mr{ISCO}}$ independent of the spin. The presence of an accretion disk truncated within a few $R_{\mr{ISCO}}$ from the black hole is consistent with the signs  of direct disk emission in the X-ray spectra. If \sw13\ is located close to us, so within 2.3 kpc, and contains a black hole with more than 4 M\subsun, we observed it at less than 0.02\% Eddington luminosity. At such low luminosities one expects to find a disk truncated far away from the black hole, as the disk is found to recede in the quiescent state \citep{1995ApJ...442..358M,2001ApJ...555..477M,2003ApJ...593..435M,1995ApJ...452..710N,1996ApJ...457..821N,1997ApJ...489..865E,2001ApJ...555..483E}. A disk truncated at 5 -- 12 $R_{\mr{ISCO}}$ has been observed at $\sim$0.02\% $L_{\mr{Edd}}$ in GRS\,1739--278 \citep{2016ApJ...832..115F}. 
Even if \sw13\ is located at $\sim$6 kpc, the luminosity is less than 0.15 per cent Eddington and the disk should be truncated far away from the black hole. In \gx339\ a disk truncated $\ga35 R_g$ has been observed at $\sim$0.14\% $L_{\mr{Edd}}$ \citep{2009ApJ...707L..87T}. Disk radii consistent with a disk extending down to the ISCO in the LHS have been reported in \gx339\ at luminosities $\ge1$\% $L_{\mr{Edd}}$ \citep{2006ApJ...653..525M,2008ApJ...680..593T,2014A&A...564A..37P}. \citet{2010MNRAS.407.2287D} argues that the small disk truncation radius observed in \citet{2006ApJ...653..525M} and \citet{2008ApJ...680..593T} is an artefact caused by severe pile-up in the \xmm\ MOS data, which broadens the iron line. As we observe \sw13\ at much lower luminosity and with \nus\ we can exclude that our spectral fits are affected by pile-up, and hence the small observed disk radius cannot be attributed to this effect.

If we assume a low inclination of 30\deg, and a spin value of $a\ge0.9$, we find an accretion disk being truncated above 4--5 $R_{\mr{ISCO}}$, consistent with the expectations for the truncation radius at such low luminosities. However, such a low inclination does not agree with the value derived from optical light curves. Regarding the inclination of \sw13, the exact geometry is still puzzling, as the dips that are observed in optical light curves and that suggest a high inclination do not show up in X-ray light curves \citep{2013Sci...339.1048C,2014MNRAS.439.3908A}. Unfortunately, the light curve of the \nus\ data used in this study is not conclusive, regarding the presence of dip-like features. 

We observed in the \nus\ cospectra a peaked noise component at $\sim$ 20 mHz, but cannot conclusively relate it to the optical dip features with show up at frequencies of $\sim$5--10 mHz in a power spectrum of SALT data, obtained simultaneously to our \nus\ observation (D. Buckley; private communication). So the variability feature in the X-rays does not seem to be a harmonic of the ones seen in the optical band. We would also like to mention that the \nus\ cospectra do not show the presence of a mHz QPO, which has been observed close to outburst peak in 2011 at a luminosity of $\sim$\oergs{35} \citep[when the source was about a factor 51.5 brighter than during the \nus\ observation; assuming a distance of 1.5 kpc;][]{2014MNRAS.439.3908A}. The \nus\ cospectra show in addition to the peaked noise component two BLN components at characteristic frequencies of $\nu=6.0^{+1.6}_{-1.1}$ mHz and $\nu=0.31^{+0.06}_{-0.05}$ Hz, respectively. Two BLN components have also been detected in RXTE PDS of the 2011 outburst, but at higher frequencies of $\nu=39\pm7$ mHz and $\nu=0.6\pm1$ Hz \citep{2014MNRAS.439.3908A}. An increase of the characteristic frequency of BLN components with increasing source luminosity has been observed in the LHS at outburst begin of (bright) low-mass X-ray binaries like \gx339\ \citep{2005A&A...440..207B}. Despite the much lower source luminosity in the \nus\ data, the rms variability of the BLN components is rather similar between the 2011 RXTE PDS \citep[$23\pm1$\% and $21\pm1$\%;][]{2014MNRAS.439.3908A} and the \nus\ cospectra ($19^{+2}_{-1}$\% and $23\pm1$\%).

\acknowledgments
We thank Fiona Harrison, Brad Cenko, and the schedulers and SOC of \swift\ and \nus\ for making these observations possible.
This project is supported by the Ministry of Science and Technology of
the Republic of China (Taiwan) through grants 105-2112-M-007-033-MY2 and 105-2811-M-007-065.
This research has made use of data obtained through the High Energy Astrophysics Science Archive Research Center Online Service, provided by the NASA/Goddard Space Flight Center.

{\it Facilities:} \facility{\nus}, \facility{\swift}.

\bibliographystyle{apj}
\bibliography{papers/my2010,papers/my2013}

\begin{thebibliography}{}
\expandafter\ifx\csname natexlab\endcsname\relax\def\natexlab#1{#1}\fi

\bibitem[{{Akaike}(1974)}]{1974ITAC...19..716A}
{Akaike}, H. 1974, IEEE Transactions on Automatic Control, 19, 716

\bibitem[{{Armas Padilla} {et~al.}(2013){Armas Padilla}, {Degenaar}, {Russell},
  \& {Wijnands}}]{2013MNRAS.428.3083A}
{Armas Padilla}, M., {Degenaar}, N., {Russell}, D.~M., \& {Wijnands}, R. 2013,
  \mnras, 428, 3083

\bibitem[{{Armas Padilla} {et~al.}(2014){Armas Padilla}, {Wijnands},
  {Altamirano}, {M{\'e}ndez}, {Miller}, \& {Degenaar}}]{2014MNRAS.439.3908A}
{Armas Padilla}, M., {Wijnands}, R., {Altamirano}, D., {et~al.} 2014, \mnras,
  439, 3908

\bibitem[{{Bachetti}(2015)}]{2015ascl.soft02021B}
{Bachetti}, M. 2015, {MaLTPyNT: Quick look timing analysis for NuSTAR data},
  Astrophysics Source Code Library, ascl:1502.021

\bibitem[{{Bachetti} {et~al.}(2015){Bachetti}, {Harrison}, {Cook}, {Tomsick},
  {Schmid}, {Grefenstette}, {Barret}, {Boggs}, {Christensen}, {Craig},
  {Fabian}, {F{\"u}rst}, {Gandhi}, {Hailey}, {Kara}, {Maccarone}, {Miller},
  {Pottschmidt}, {Stern}, {Uttley}, {Walton}, {Wilms}, \&
  {Zhang}}]{2015ApJ...800..109B}
{Bachetti}, M., {Harrison}, F.~A., {Cook}, R., {et~al.} 2015, \apj, 800, 109

\bibitem[{{Barthelmy} {et~al.}(2005){Barthelmy}, {Barbier}, {Cummings},
  {Fenimore}, {Gehrels}, {Hullinger}, {Krimm}, {Markwardt}, {Palmer},
  {Parsons}, {Sato}, {Suzuki}, {Takahashi}, {Tashiro}, \&
  {Tueller}}]{2005SSRv..120..143B}
{Barthelmy}, S.~D., {Barbier}, L.~M., {Cummings}, J.~R., {et~al.} 2005, \ssr,
  120, 143

\bibitem[{{Basak} \& {Zdziarski}(2016)}]{2016MNRAS.458.2199B}
{Basak}, R., \& {Zdziarski}, A.~A. 2016, \mnras, 458, 2199

\bibitem[{{Belloni} {et~al.}(2005){Belloni}, {Homan}, {Casella}, {van der
  Klis}, {Nespoli}, {Lewin}, {Miller}, \& {M{\'e}ndez}}]{2005A&A...440..207B}
{Belloni}, T., {Homan}, J., {Casella}, P., {et~al.} 2005, \aap, 440, 207

\bibitem[{{Burnham} {et~al.}(2011){Burnham}, {Anderson}, \&
  {Huyvaert}}]{burnham2011}
{Burnham}, K.~P., {Anderson}, D.~R., \& {Huyvaert}, K.~P. 2011, Behav. Ecol.
  Sociobiol., 65, 23

\bibitem[{{Burrows} {et~al.}(2005){Burrows}, {Hill}, {Nousek}, {Kennea},
  {Wells}, {Osborne}, {Abbey}, {Beardmore}, {Mukerjee}, {Short}, {Chincarini},
  {Campana}, {Citterio}, {Moretti}, {Pagani}, {Tagliaferri}, {Giommi},
  {Capalbi}, {Tamburelli}, {Angelini}, {Cusumano}, {Br{\"a}uninger}, {Burkert},
  \& {Hartner}}]{2005SSRv..120..165B}
{Burrows}, D.~N., {Hill}, J.~E., {Nousek}, J.~A., {et~al.} 2005, \ssr, 120, 165

\bibitem[{{Cardelli} {et~al.}(1989){Cardelli}, {Clayton}, \&
  {Mathis}}]{1989ApJ...345..245C}
{Cardelli}, J.~A., {Clayton}, G.~C., \& {Mathis}, J.~S. 1989, \apj, 345, 245

\bibitem[{{Corral-Santana} {et~al.}(2013){Corral-Santana}, {Casares},
  {Mu{\~n}oz-Darias}, {Rodr{\'{\i}}guez-Gil}, {Shahbaz}, {Torres}, {Zurita}, \&
  {Tyndall}}]{2013Sci...339.1048C}
{Corral-Santana}, J.~M., {Casares}, J., {Mu{\~n}oz-Darias}, T., {et~al.} 2013,
  Science, 339, 1048

\bibitem[{{Dauser} {et~al.}(2014){Dauser}, {Garc{\'{\i}}a}, {Parker}, {Fabian},
  \& {Wilms}}]{2014MNRAS.444L.100D}
{Dauser}, T., {Garc{\'{\i}}a}, J., {Parker}, M.~L., {Fabian}, A.~C., \&
  {Wilms}, J. 2014, \mnras, 444, L100

\bibitem[{{Done} \& {Diaz Trigo}(2010)}]{2010MNRAS.407.2287D}
{Done}, C., \& {Diaz Trigo}, M. 2010, \mnras, 407, 2287

\bibitem[{{Drake} {et~al.}(2017){Drake}, {Djorgovski}, {Mahabal}, {Graham},
  {Stern}, {Catelan}, {Christensen}, \& {Larson}}]{2017ATel10297....1D}
{Drake}, A.~J., {Djorgovski}, S.~G., {Mahabal}, A.~A., {et~al.} 2017, The
  Astronomer's Telegram, No.~10297, 297

\bibitem[{{Esin} {et~al.}(2001){Esin}, {McClintock}, {Drake}, {Garcia},
  {Haswell}, {Hynes}, \& {Muno}}]{2001ApJ...555..483E}
{Esin}, A.~A., {McClintock}, J.~E., {Drake}, J.~J., {et~al.} 2001, \apj, 555,
  483

\bibitem[{{Esin} {et~al.}(1997){Esin}, {McClintock}, \&
  {Narayan}}]{1997ApJ...489..865E}
{Esin}, A.~A., {McClintock}, J.~E., \& {Narayan}, R. 1997, \apj, 489, 865

\bibitem[{{Evans} {et~al.}(2009){Evans}, {Beardmore}, {Page}, {Osborne},
  {O'Brien}, {Willingale}, {Starling}, {Burrows}, {Godet}, {Vetere}, {Racusin},
  {Goad}, {Wiersema}, {Angelini}, {Capalbi}, {Chincarini}, {Gehrels}, {Kennea},
  {Margutti}, {Morris}, {Mountford}, {Pagani}, {Perri}, {Romano}, \&
  {Tanvir}}]{2009MNRAS.397.1177E}
{Evans}, P.~A., {Beardmore}, A.~P., {Page}, K.~L., {et~al.} 2009, \mnras, 397,
  1177

\bibitem[{{F{\"u}rst} {et~al.}(2016){F{\"u}rst}, {Tomsick}, {Yamaoka},
  {Dauser}, {Miller}, {Clavel}, {Corbel}, {Fabian}, {Garc{\'{\i}}a},
  {Harrison}, {Loh}, {Kaaret}, {Kalemci}, {Migliari}, {Miller-Jones},
  {Pottschmidt}, {Rahoui}, {Rodriguez}, {Stern}, {Stuhlinger}, {Walton}, \&
  {Wilms}}]{2016ApJ...832..115F}
{F{\"u}rst}, F., {Tomsick}, J.~A., {Yamaoka}, K., {et~al.} 2016, \apj, 832, 115

\bibitem[{{Garc{\'{\i}}a} {et~al.}(2014){Garc{\'{\i}}a}, {Dauser}, {Lohfink},
  {Kallman}, {Steiner}, {McClintock}, {Brenneman}, {Wilms}, {Eikmann},
  {Reynolds}, \& {Tombesi}}]{2014ApJ...782...76G}
{Garc{\'{\i}}a}, J., {Dauser}, T., {Lohfink}, A., {et~al.} 2014, \apj, 782, 76

\bibitem[{{Gierli{\'n}ski} {et~al.}(2008){Gierli{\'n}ski}, {Done}, \&
  {Page}}]{2008MNRAS.388..753G}
{Gierli{\'n}ski}, M., {Done}, C., \& {Page}, K. 2008, \mnras, 388, 753

\bibitem[{{Gierli{\'n}ski} {et~al.}(2009){Gierli{\'n}ski}, {Done}, \&
  {Page}}]{2009MNRAS.392.1106G}
---. 2009, \mnras, 392, 1106

\bibitem[{{Harrison} {et~al.}(2013){Harrison}, {Craig}, {Christensen},
  {Hailey}, {Zhang}, {Boggs}, {Stern}, {Cook}, {Forster}, {Giommi},
  {Grefenstette}, {Kim}, {Kitaguchi}, {Koglin}, {Madsen}, {Mao}, {Miyasaka},
  {Mori}, {Perri}, {Pivovaroff}, {Puccetti}, {Rana}, {Westergaard}, {Willis},
  {Zoglauer}, {An}, {Bachetti}, {Barri{\`e}re}, {Bellm}, {Bhalerao},
  {Brejnholt}, {Fuerst}, {Liebe}, {Markwardt}, {Nynka}, {Vogel}, {Walton},
  {Wik}, {Alexander}, {Cominsky}, {Hornschemeier}, {Hornstrup}, {Kaspi},
  {Madejski}, {Matt}, {Molendi}, {Smith}, {Tomsick}, {Ajello}, {Ballantyne},
  {Balokovi{\'c}}, {Barret}, {Bauer}, {Blandford}, {Brandt}, {Brenneman},
  {Chiang}, {Chakrabarty}, {Chenevez}, {Comastri}, {Dufour}, {Elvis}, {Fabian},
  {Farrah}, {Fryer}, {Gotthelf}, {Grindlay}, {Helfand}, {Krivonos}, {Meier},
  {Miller}, {Natalucci}, {Ogle}, {Ofek}, {Ptak}, {Reynolds}, {Rigby},
  {Tagliaferri}, {Thorsett}, {Treister}, \& {Urry}}]{2013ApJ...770..103H}
{Harrison}, F.~A., {Craig}, W.~W., {Christensen}, F.~E., {et~al.} 2013, \apj,
  770, 103

\bibitem[{{Houck} \& {Denicola}(2000)}]{2000ASPC..216..591H}
{Houck}, J.~C., \& {Denicola}, L.~A. 2000, in Astronomical Society of the
  Pacific Conference Series, Vol. 216, Astronomical Data Analysis Software and
  Systems IX, ed. {N.~Manset, C.~Veillet, \& D.~Crabtree}, 591

\bibitem[{{Krimm} {et~al.}(2011){Krimm}, {Barthelmy}, {Baumgartner},
  {Cummings}, {Fenimore}, {Gehrels}, {Markwardt}, {Palmer}, {Sakamoto},
  {Skinner}, {Stamatikos}, {Tueller}, \& {Ukwatta}}]{2011ATel.3138....1K}
{Krimm}, H.~A., {Barthelmy}, S.~D., {Baumgartner}, W., {et~al.} 2011, The
  Astronomer's Telegram, 3138

\bibitem[{{Mata S{\'a}nchez} {et~al.}(2015){Mata S{\'a}nchez},
  {Mu{\~n}oz-Darias}, {Casares}, {Corral-Santana}, \&
  {Shahbaz}}]{2015MNRAS.454.2199M}
{Mata S{\'a}nchez}, D., {Mu{\~n}oz-Darias}, T., {Casares}, J.,
  {Corral-Santana}, J.~M., \& {Shahbaz}, T. 2015, \mnras, 454, 2199

\bibitem[{{McClintock} {et~al.}(1995){McClintock}, {Horne}, \&
  {Remillard}}]{1995ApJ...442..358M}
{McClintock}, J.~E., {Horne}, K., \& {Remillard}, R.~A. 1995, \apj, 442, 358

\bibitem[{{McClintock} {et~al.}(2003){McClintock}, {Narayan}, {Garcia},
  {Orosz}, {Remillard}, \& {Murray}}]{2003ApJ...593..435M}
{McClintock}, J.~E., {Narayan}, R., {Garcia}, M.~R., {et~al.} 2003, \apj, 593,
  435

\bibitem[{{McClintock} {et~al.}(2001){McClintock}, {Haswell}, {Garcia},
  {Drake}, {Hynes}, {Marshall}, {Muno}, {Chaty}, {Garnavich}, {Groot}, {Lewin},
  {Mauche}, {Miller}, {Pooley}, {Shrader}, \& {Vrtilek}}]{2001ApJ...555..477M}
{McClintock}, J.~E., {Haswell}, C.~A., {Garcia}, M.~R., {et~al.} 2001, \apj,
  555, 477

\bibitem[{{Miller} {et~al.}(2006){Miller}, {Homan}, {Steeghs}, {Rupen},
  {Hunstead}, {Wijnands}, {Charles}, \& {Fabian}}]{2006ApJ...653..525M}
{Miller}, J.~M., {Homan}, J., {Steeghs}, D., {et~al.} 2006, \apj, 653, 525

\bibitem[{{Narayan} {et~al.}(1996){Narayan}, {McClintock}, \&
  {Yi}}]{1996ApJ...457..821N}
{Narayan}, R., {McClintock}, J.~E., \& {Yi}, I. 1996, \apj, 457, 821

\bibitem[{{Narayan} \& {Yi}(1995)}]{1995ApJ...452..710N}
{Narayan}, R., \& {Yi}, I. 1995, \apj, 452, 710

\bibitem[{{Petrucci} {et~al.}(2014){Petrucci}, {Cabanac}, {Corbel}, {Koerding},
  \& {Fender}}]{2014A&A...564A..37P}
{Petrucci}, P.-O., {Cabanac}, C., {Corbel}, S., {Koerding}, E., \& {Fender}, R.
  2014, \aap, 564, A37

\bibitem[{{Plotkin} {et~al.}(2016){Plotkin}, {Gallo}, {Jonker}, {Miller-Jones},
  {Homan}, {Mu{\~n}oz-Darias}, {Markoff}, {Armas Padilla}, {Fender}, {Rushton},
  {Russell}, \& {Torres}}]{2016MNRAS.456.2707P}
{Plotkin}, R.~M., {Gallo}, E., {Jonker}, P.~G., {et~al.} 2016, \mnras, 456,
  2707

\bibitem[{{Rau} {et~al.}(2011){Rau}, {Greiner}, \&
  {Filgas}}]{2011ATel.3140....1R}
{Rau}, A., {Greiner}, J., \& {Filgas}, R. 2011, The Astronomer's Telegram, 3140

\bibitem[{{Reynolds} \& {Begelman}(1997)}]{1997ApJ...488..109R}
{Reynolds}, C.~S., \& {Begelman}, M.~C. 1997, \apj, 488, 109

\bibitem[{{Roming} {et~al.}(2005){Roming}, {Kennedy}, {Mason}, {Nousek}, {Ahr},
  {Bingham}, {Broos}, {Carter}, {Hancock}, {Huckle}, {Hunsberger}, {Kawakami},
  {Killough}, {Koch}, {McLelland}, {Smith}, {Smith}, {Soto}, {Boyd},
  {Breeveld}, {Holland}, {Ivanushkina}, {Pryzby}, {Still}, \&
  {Stock}}]{2005SSRv..120...95R}
{Roming}, P.~W.~A., {Kennedy}, T.~E., {Mason}, K.~O., {et~al.} 2005, \ssr, 120,
  95

\bibitem[{{Schlegel} {et~al.}(1998){Schlegel}, {Finkbeiner}, \&
  {Davis}}]{1998ApJ...500..525S}
{Schlegel}, D.~J., {Finkbeiner}, D.~P., \& {Davis}, M. 1998, \apj, 500, 525

\bibitem[{{Shahbaz} {et~al.}(2013){Shahbaz}, {Russell}, {Zurita}, {Casares},
  {Corral-Santana}, {Dhillon}, \& {Marsh}}]{2013MNRAS.434.2696S}
{Shahbaz}, T., {Russell}, D.~M., {Zurita}, C., {et~al.} 2013, \mnras, 434, 2696

\bibitem[{{Sivakoff} {et~al.}(2017){Sivakoff}, {Tetarenko}, {Shaw}, \&
  {Bahramian}}]{2017ATel10314....1S}
{Sivakoff}, G.~R., {Tetarenko}, B.~E., {Shaw}, A.~W., \& {Bahramian}, A. 2017,
  The Astronomer's Telegram, No.~10314, 314

\bibitem[{{Stiele} \& {Kong}(2017)}]{2017ApJ...844....8S}
{Stiele}, H., \& {Kong}, A.~K.~H. 2017, \apj, 844, 8

\bibitem[{{Tomsick} {et~al.}(2009){Tomsick}, {Yamaoka}, {Corbel}, {Kaaret},
  {Kalemci}, \& {Migliari}}]{2009ApJ...707L..87T}
{Tomsick}, J.~A., {Yamaoka}, K., {Corbel}, S., {et~al.} 2009, \apjl, 707, L87

\bibitem[{{Tomsick} {et~al.}(2008){Tomsick}, {Kalemci}, {Kaaret}, {Markoff},
  {Corbel}, {Migliari}, {Fender}, {Bailyn}, \& {Buxton}}]{2008ApJ...680..593T}
{Tomsick}, J.~A., {Kalemci}, E., {Kaaret}, P., {et~al.} 2008, \apj, 680, 593

\bibitem[{{Torres} {et~al.}(2015){Torres}, {Jonker}, {Miller-Jones}, {Steeghs},
  {Repetto}, \& {Wu}}]{2015MNRAS.450.4292T}
{Torres}, M.~A.~P., {Jonker}, P.~G., {Miller-Jones}, J.~C.~A., {et~al.} 2015,
  \mnras, 450, 4292

\bibitem[{{Verner} {et~al.}(1996){Verner}, {Ferland}, {Korista}, \&
  {Yakovlev}}]{1996ApJ...465..487V}
{Verner}, D.~A., {Ferland}, G.~J., {Korista}, K.~T., \& {Yakovlev}, D.~G. 1996,
  \apj, 465, 487

\bibitem[{{Wijnands} {et~al.}(2004){Wijnands}, {Homan}, {Miller}, \&
  {Lewin}}]{2004ApJ...606L..61W}
{Wijnands}, R., {Homan}, J., {Miller}, J.~M., \& {Lewin}, W.~H.~G. 2004, \apjl,
  606, L61

\bibitem[{{Wijnands} {et~al.}(2006){Wijnands}, {in't Zand}, {Rupen},
  {Maccarone}, {Homan}, {Cornelisse}, {Fender}, {Grindlay}, {van der Klis},
  {Kuulkers}, {Markwardt}, {Miller-Jones}, \& {Wang}}]{2006A&A...449.1117W}
{Wijnands}, R., {in't Zand}, J.~J.~M., {Rupen}, M., {et~al.} 2006, \aap, 449,
  1117

\bibitem[{{Wilms} {et~al.}(2000){Wilms}, {Allen}, \&
  {McCray}}]{2000ApJ...542..914W}
{Wilms}, J., {Allen}, A., \& {McCray}, R. 2000, \apj, 542, 914

\bibitem[{{Zdziarski} {et~al.}(1996){Zdziarski}, {Johnson}, \&
  {Magdziarz}}]{1996MNRAS.283..193Z}
{Zdziarski}, A.~A., {Johnson}, W.~N., \& {Magdziarz}, P. 1996, \mnras, 283, 193

\bibitem[{{{\.Z}ycki} {et~al.}(1999){{\.Z}ycki}, {Done}, \&
  {Smith}}]{1999MNRAS.309..561Z}
{{\.Z}ycki}, P.~T., {Done}, C., \& {Smith}, D.~A. 1999, \mnras, 309, 561

\end{thebibliography}

%\appendix

%\section[]{Online Material}
%\bsp

%\label{lastpage}

\end{document}